\documentclass[preprint,amsmath,amssymb,aps,showpacs,showkeys]{revtex4}
\usepackage{graphicx}
\usepackage{graphics}
\usepackage{amsmath}
\usepackage{dcolumn}
\usepackage{bm}
\usepackage{xcolor}
\usepackage{amsmath}
\usepackage{amssymb}
\begin{document}
%%%%%%%%%%%%%%%%%%%%%%%%%%%%%%%%%%%%%%%%%%%%%%%%%%%%%%%%%%%%%%%%%%%%%%
%%%%%%%%%%%%%%%%%%%%%%%%%%%%%%%%%%%%%%%%%%%%%%%%%%%%%%%%%%%%%%%%%%%%%%
\title{Scaterring of Massless Quasiparticles in the $^{3}He-A$ Superfluid}
%%%%%%%%%%%%%%%%%%%%%%%%%%%%%%%%%%%%%%%%%%%%%%%%%%%%%%%%%%%%%%%%%%%%%%
%%%%%%%%%%%%%%%%%%%%%%%%%%%%%%%%%%%%%%%%%%%%%%%%%%%%%%%%%%%%%%%%%%%%%%
\author{Alexandre M. de M. Carvalho\footnote{alexandre@fis.ufal.br}}
%%%%%%%%%%%%%%%%%%%%%%%%%%%%%%%%%%%%%%%%%%%%%%%%%%%%%%%%%%%%%%%%%%%%%%
\affiliation{Instituto de F\'{\i}sica, Universidade Federal de Alagoas, Campus A. C. Sim\~oes - Av. Lourival Melo Mota, s/n, Tabuleiro do Martins, 57072-970, Macei\'o,  AL, Brazil}
%%%%%%%%%%%%%%%%%%%%%%%%%%%%%%%%%%%%%%%%%%%%%%%%%%%%%%%%%%%%%%%%%%%%%%
\author{~Alisson S. Marques\footnote{alisson.marques@fis.ufal.br}}
%%%%%%%%%%%%%%%%%%%%%%%%%%%%%%%%%%%%%%%%%%%%%%%%%%%%%%%%%%%%%%%%%%%%%%
\affiliation{Instituto de F\'{\i}sica, Universidade Federal de Alagoas, Campus A. C. Sim\~oes - Av. Lourival Melo Mota, s/n, Tabuleiro do Martins, 57072-970, Macei\'o,  AL, Brazil}
%%%%%%%%%%%%%%%%%%%%%%%%%%%%%%%%%%%%%%%%%%%%%%%%%%%%%%%%%%%%%%%%%%%%%%
\author{Glauber T. Silva\footnote{glauber@fis.ufal.br}}
%%%%%%%%%%%%%%%%%%%%%%%%%%%%%%%%%%%%%%%%%%%%%%%%%%%%%%%%%%%%%%%%%%%%%%
\affiliation{Instituto de F\'{\i}sica, Universidade Federal de Alagoas, Campus A. C. Sim\~oes - Av. Lourival Melo Mota, s/n, Tabuleiro do Martins, 57072-970, Macei\'o,  AL, Brazil}
%%%%%%%%%%%%%%%%%%%%%%%%%%%%%%%%%%%%%%%%%%%%%%%%%%%%%%%%%%%%%%%%%%%%%%
\author{Gabriel Q. Garcia\footnote{gqgarcia99@gmail.com}}
%%%%%%%%%%%%%%%%%%%%%%%%%%%%%%%%%%%%%%%%%%%%%%%%%%%%%%%%%%%%%%%%%%%%%%
\affiliation{Centro de Ci\^encias, Tecnologia e Sa\'ude, Universidade Estadual da Para\'iba, 58233-000, Araruna, PB, Brazil.}
%%%%%%%%%%%%%%%%%%%%%%%%%%%%%%%%%%%%%%%%%%%%%%%%%%%%%%%%%%%%%%%%%%%%%%
\author{Claudio Furtado\footnote{furtado@fisica.ufpb.br}}
%%%%%%%%%%%%%%%%%%%%%%%%%%%%%%%%%%%%%%%%%%%%%%%%%%%%%%%%%%%%%%%%%%%%%%
\affiliation{Departamento de F\'{\i}sica, CCEN,  Universidade Federal da Para\'{\i}ba, Cidade Universit\'{a}ria, 58051-970, Jo\~ao Pessoa, PB, Brazil.}
%%%%%%%%%%%%%%%%%%%%%%%%%%%%%%%%%%%%%%%%%%%%%%%%%%%%%%%%%%%%%%%%%%%%%%
%%%%%%%%%%%%%%%%%%%%%%%%%%%%%%%%%%%%%%%%%%%%%%%%%%%%%%%%%%%%%%%%%%%%%%
\begin{abstract}
In this work, we analyze the scattering of fermionic quasiparticles in the presence of radial disgyrations and symmetric vortices in the superfluid \( ^3\text{He-A} \). We consider a Volovik analog model for the description of these defects and investigate the scattering of fermionic quasiparticles in this background. Furthermore, we solve the massless Dirac equation employing this methodology to gain comprehensive insights into the scattering phenomena and its dependence on the geometric properties.  These results validate the optical theorem and highlight the role of defect topology in the scattering process.
\end{abstract}
\maketitle

%%%%%%%%%%%%%%%%%%%%%%%%%%%%%%%%%%%%%%%%%%%%%%%%%%%%%%%%%%%%%%%%%%%%%%
%%%%%%%%%%%%%%%%%%%%%%%%%%%%%%%%%%%%%%%%%%%%%%%%%%%%%%%%%%%%%%%%%%%%%%
\section{Introduction}
%%%%%%%%%%%%%%%%%%%%%%%%%%%%%%%%%%%%%%%%%%%%%%%%%%%%%%%%%%%%%%%%%%%%%%
%%%%%%%%%%%%%%%%%%%%%%%%%%%%%%%%%%%%%%%%%%%%%%%%%%%%%%%%%%%%%%%%%%%%%%
During the last few years condensed matter systems whose behavior is analogous to general relativity systems were intensively studied. The analogy between condensed matter and gravity is so strong that several recent cosmological experiments have been replayed with liquid crystals \cite{scie:bow} and superfluid helium \cite{pr:zur}, shading a new light on the problem of dynamics of defect formations. This interdisciplinarity has attracted strong attention to condensed matter systems which can be used as a laboratory for studying gravitational systems. These condensed matter systems allowing us to simulate gravitational phenomena were denominated as analog models.

 Various condensed matter systems have been treated as analog models. Among many, we can emphasize the following ones: Bose-Einstein condensates ~\cite{prl:gar,pra:gar}, classical fluids~\cite{prl:unruh,prd:unruh,prd:jac,prl:visser,cqg:visser} and quantum fluids ~\cite{volo}-\cite{boo}, moving dielectric media~\cite{leon,brev} and non-linear electrodynamics~\cite{prd:novello}. Another interesting research in this context is the study of non-commutative acoustic black holes in a non-commutative Abelian Higgs model~\cite{eduardo}, where it was shown that the Abelian Higgs model is efficient for application to high-energy physics, and the non-commutative Abelian Higgs model can also describe Lorentz-symmetry violation in high-energy particle physics.  The acoustic superresonance phenomenon in acoustic black holes, that is, the analogue to the superradiance in black hole physics, was studied in Refs. \cite{anacleto1,anacleto2}. Recently, the non-commutative analog of the Aharonov-Bohm effect for an acoustic vortex was considered in Ref. \cite{anacleto3}
 
Our aim in this paper is to investigate the classical and quantum dynamics of a particle in spacetime with conical singularities in (3+1)-dimensional space. The problem of particle scattering in the presence of these singularities was well studied in the literature in both (2 + 1)-dimensions and (3 + 1)-dimensions. Now, we intend to study some cases of important limits which can be experimentally verified in an analog condensed matter system. In this paper, we investigate the dynamics of quasiparticles within the Volovik analog model for quantum fluids. Earlier in~\cite{3} a radial disgyration described as a cosmic string was considered, which would represent a topologically stable linear defect in $^{3}He-A$. In a previous work~\cite{garciadis}, we studied the analogous Aharonov-Bohm effect in the background given by a disgyration of phonons with the use of holonomy transformations. In this case the Aharonov-Bohm effect is a global effect taking place due to topological properties of the conical spacetime. 

\section{Superfluid $^{3}He-A$ as an analogous gravitational model}

Topological defects in space-time can be characterized through metrics with zero Riemann-Chistoffel curvature tensor everywhere, except for the defects, i.e., by conic-type curvature singularities~\cite{Sta}. Typical examples of such topological defects are the domain wall~\cite{Vil1}, the cosmic string~\cite{Vil1,His} and the global monopole~\cite{Bar}. The cosmic strings provide a bridge between microscopic and macroscopic physical phenomena. They represent themselves as linear defects analogous to the vortex filaments in superfluid helium~\cite{Vol} and to dislocations and disclinations in condensed matter physics~\cite{Kat}. The cosmic string probably have been formed in the very beginning of the universe~\cite{Kib}. An infinite thin cosmic string is described in cylindrical  coordinates by the following line element:
\begin{eqnarray}
\label{string}
ds^{2}=-dt^{2}+dz^{2}+d\rho^{2}+\alpha^{2}\rho^{2}d\phi^{2},
\end{eqnarray}
where $\alpha$ is a parameter related to the linear mass density $\mu$ of the cosmic string as $\alpha=1-4G\mu$. For the cosmic string, the parameter $\alpha$ assumes only values smaller than 1. The values of parameter $\alpha$ near zero correspond to a large mass density, and this limiting case is the supermassive cosmic string. Values for $\alpha$ greater than 1,  correspond to an anticonical space-time with a negative curvature.  In a gravitational context this situation can be treated as a "``antigravitating" string. However,  it is not physically accepted in this context, but in condensed matter this ``antigravitating" string can be realized in forms of defects in solids and of topological defects in superfluid helium.

In~\cite{boo} it was showed that quasiparticles in the $^{3}He-A$ are chiral massless fermions, and their energy spectrum is given by the equation
\begin{equation}
E^{2}({\bf k})+g^{ik}(k_{i}-eA_{i})(k_{k}-eA_{k})=0
\end{equation}
where ${\bf k}$ is the wave vector and ${\bf A}$ is the vector potential which can be written as ${\bf A}=k_{F}{\bf l}$, where ${\bf l}$ is an unitary vector in the direction of the gap nodes in the momentum space, and $k_{F}$ is the Fermi momentum. This unitary vector introduces the uniaxial anisotropy of the metric tensor that describes the effective geometry that determines the dynamics of fermions.

The similarity between disgyrations and cosmic strings goes beyond their topology: for some applications, both kinds of defects can be treated through some geometrical methods. According to ~\cite{boo}, there are two types of defects: the first is the radial disgyration and the second is the axial symmetric vortex. Both a disgyration and a vortex are linear objects with non-zero winding numbers, and they can be constrained to the same plane. Other interesting property shared by disgyrations and vortices is that the degeneracy parameters are well defined at the singularity ${\rho}=0$, except for the London energy diverging at the origin. For the radial disgyration the elastic energy is infinite at ${\rho}=0$, while for the quantized vortex the kinetic energy of the superflow is also infinite at ${\rho}=0$. Such singularities can be studied by topological methods.

The radial disgyrations are characterized by singularities of $\vec{l}$-vector field, it is represented by the following set of basic vectors:
\begin{eqnarray}\label{lrad}
\hat{e}_{1}=\hat{\phi}, \quad \quad \hat{e}_{2}=\hat{z},\quad \quad \hat{l}=\hat{\rho},
\end{eqnarray}
while for axial case the distribution field configurations $\vec{l}$ are represented by
\begin{eqnarray}\label{laxi}
\hat{e}_{1}=\hat{z}, \quad \quad \hat{e}_{2}=\hat{\rho},\quad \quad \hat{l}=\hat{\phi}.
\end{eqnarray}
Here and further, we will consider both defects, radial disgyrations and the symmetric vortex, sharing the above properties with vortices being therefore analogous to cosmic strings. In the Volovik analogous model the effective  metrics for the quasiparticles moving outside of the core of the radial and axial disgyrations are described by the following metric tensor:
\begin{eqnarray}
g^{00} = -1,\ g^{i0} = v^{i}_{s}\ \text{and}\ g^{ik} = c^{2}_{||} l^{i} l^{k}+ c^{2}_{\bot}(\delta^{ik}- l^{i} l^{k}) - v^{i}_{s} v^{i}_{s},
\end{eqnarray}
Note that the vector $\vec{v}_{s}$ is playing a role of the superfluid velocity. Thus, the disgyration provides only the ``gravity" field which acts on the $^{3}He-A$ fermions. In deal with the conditions \eqref{lrad} and considering the null velocity term, the line element of the space-time of a radial disgyration is given by
\begin{equation}
ds^{2}=-dt^{2}+\frac{dz^{2}}{c^{2}_{\bot}} + \frac{1}{c^{2}_{||}} \left[d\rho^{2} + \frac{c^{2}_{||}}{c^{2}_{\bot}}\rho^{2} d\phi^{2}\right].
\label{LErad}
\end{equation}
For the symmetric vortex, we consider the conditions \eqref{laxi} with the superfluid velocity given by $\vec{v}_s = \frac{\hslash}{2m_3 \rho}\hat{\phi}$. And then, we can describe the space-time geometry through of the line element gives bellow
\begin{eqnarray}
ds^{2}=-\left(1 - \frac{v^{2}_{s}}{c^{2}_{\bot}}\right)\left(dt + \frac{\hslash d\phi}{2m_3(c^{2}_{\bot} - v^{2}_{s})}\right)^2 + \frac{dz^2}{c^{2}_{||}} + \frac{1}{c^{2}_{\bot}} \left[d\rho^{2} + \frac{c^{2}_{\bot} \rho^{2} d\phi^{2}}{(c^{2}_{\bot} - v^{2}_{s})}\right],
\label{LEaxi}
\end{eqnarray}
where $c_{\bot}$ and $c_{||}$ represent respectively the ``speeds of light" orthogonal and parallel to the gap nodes direction in the momentum space. An interesting case occurs when the quasiparticles are far from the vortex axis, so we can ignore the superfluid velocity term $\vec{v}_{s}$. Therefore, considering the angular velocity as $\omega = \frac{2m_3 c^{2}_{\bot}}{\hslash}$, the line element \eqref{LEaxi} assumes the following form:
\begin{eqnarray}
    ds^2 = -\left(dt + \frac{d\phi}{\omega}\right)^2 + \frac{dz^2}{c^{2}_{||}} + \frac{1}{c^{2}_{\bot}} \left[d\rho^{2} + \rho^{2} d\phi^{2}\right].
    \label{LEaxi2}
\end{eqnarray}

%%%%%%%%%%%%%%%%%%%%%%%%%%%%%%%%%%%%%%%%%%%%%%%%%%%%%%%%%%%%%%%%%%%%%%
%%%%%%%%%%%%%%%%%%%%%%%%%%%%%%%%%%%%%%%%%%%%%%%%%%%%%%%%%%%%%%%%%%%%%%
\section{Scattering by a Radial Disgyration}
%%%%%%%%%%%%%%%%%%%%%%%%%%%%%%%%%%%%%%%%%%%%%%%%%%%%%%%%%%%%%%%%%%%%%%
%%%%%%%%%%%%%%%%%%%%%%%%%%%%%%%%%%%%%%%%%%%%%%%%%%%%%%%%%%%%%%%%%%%%%%

 Let us begin with the case of radial dysgiration. From the line element \eqref{LErad}, we can note that this geometry possesses a conical singularity if $c_{||}$ does not coincide with $c_{\bot}$. This conical singularity is represented by the following curvature tensor
\begin{eqnarray}
\label{curv}
R_{\rho,\phi}^{\rho,\phi} = \frac{1-\alpha_{r}}{4G\alpha_{r}}\delta_{2}(\vec{\rho}),
\end{eqnarray}
Here $\delta_{2}(\vec{\rho})$ is a two-dimensional delta function and $\alpha_{r}= c_{||}/c_{\bot}$. This term is identified as the excess or deficit angle in the cosmic string. This behavior of the curvature tensor is called the conical singularity ~\cite{staro}, which gives rise to the curvature concentrated on the disgyration axis. This metric pointed out in \cite{volo} is similar to the cosmic string. A crucial difference is the fact that the radial disgyration metric has $\alpha^{2}>1$. We also note that an analogous situation occurs for defects in solids~\cite{Kro,kat}. For simplicity, we can introduce a coordinate transformation in equation (\ref{LErad}), 
\begin{subequations}
\begin{eqnarray}
z &\rightarrow& c_{\bot} z';\\
\rho &\rightarrow& c_{||} \rho';\\
t & \rightarrow& c_{||} t'.
\end{eqnarray}
\end{subequations}
And then we can rewrite the line element \eqref{LErad}
\begin{eqnarray}
\label{metrica_efetiva}
ds^{2}=-c^{2}_{||}dt^{2}+d\rho^{2}+ \alpha_{r}^{2} \rho^{2}d\phi^{2} +dz^{2},
\end{eqnarray}
in order to simplify the notation, we omitted the prime index at the coordinates.

Now, we are interested in obtaining the scattering amplitude and the total cross section for the scattering of fermionic quasiparticles in the presence of a disgyration in a superfluid within a curved space description. We use the formalism developed by Adhikari~\cite{adhikari} to obtain the scattering amplitude and the optical theorem. This approach was also used by Azevedo and Moraes~\cite{azevedo} in order to describe the quantum scattering by disclination in a graphite sheet in a non-relativistic regime. As was pointed out in~\cite{volo}, the fermions in the A-phase near the gap nodes are equivalent to the relativistic massless charged particles with electric charge $e$, moving in electromagnetic and gravitational fields~\cite{exotic}. The  Dirac equation for a massless particle on a curved space-time is given by
\begin{equation}
\label{dirac}
i\gamma^{a}e^{\ \mu}_{a}(x) D_{\mu} \psi = 0,
\end{equation}
where  $\psi$ is a two-component spinor and $D_{\mu}$ is the covariant derivative  of a spinor, defined in terms of the spin connection,
\begin{equation}
D_{\mu}=\partial_{\mu}+ \frac{1}{2}\omega_{\mu ab}\sigma^{ab}.
\end{equation}
The  metric (\ref{metrica_efetiva}) allow us to choose the {\it dreibein} in $(t,\rho,\phi)$ coordinates to be
\begin{subequations}
\begin{eqnarray}
e^{\ \mu}_{0}&=&\delta^{\ \mu}_{0};\\
e^{\ \mu}_{1}&=&\cos \phi \delta^{\ \mu}_{1} - \frac{1}{\alpha_{r}}
\sin \phi \delta^{\ \mu}_{2};\\
e^{\ \mu}_{2}&=&\sin \phi \delta^{\ \mu}_{1} + \frac{1}{\alpha_{r}}
\cos \phi \delta^{\ \mu}_{2}   \mbox{.}
\end{eqnarray}
\end{subequations}
The  Dirac matrices in a curved background in terms of {\it dreibein}  are defined as
\begin{equation}
\tilde{\gamma}^{\mu}=\gamma^{a}e^{\ \mu}_{a},
\end{equation}
where $\gamma^{a}$ is the Gamma matrices in the Minkowski space-time. For metric (\ref{metrica_efetiva}), the  Dirac equation can be written  as
\begin{equation}
\label{dirac-x}
i\gamma^{0}\frac{\partial\psi}{\partial t} + i\gamma^{\rho} \left( \frac{\partial}{\partial \rho} - \frac{1-\alpha_{r}}{\alpha_{r} \rho}\right)\psi
+ i\frac{\gamma^{\phi}}{\alpha_{r} \rho} \frac{\partial\psi}{\partial \phi} = 0.
\end{equation}
Due  to the symmetry of  the equation (\ref{dirac-x}), we can choose a  positive energy solutions  as angular momentum eigenfunctions corresponding to the eigenvalue $n+1/2$:
\begin{eqnarray}
u_{n}({\bf r})e^{iEt}=\exp\{i(n+\frac{1}{2}-\frac{1}{2}\sigma^{3})\phi\}
%\nonumber \\
%\times
\left(
\begin{array}{c}
\sqrt{E}\quad u_{n}^{(1)}(\rho)  \\
\sqrt{E}\quad u_{n}^{(2)}(\rho)
\end{array}
\right)
e^{iEt},
\end{eqnarray}
where $n$ is an integer. When we substitute the explicit form of $\sigma^{3}$ in the above equation, it is reduced to the following system of equations:
\begin{eqnarray}
\left(
\begin{array}{cc}
\sqrt{E} &  i\left[\alpha_{r}\left(
\partial_{\rho}+\frac{1}{\rho}\right)+\frac{n+1/2}{\rho}\right] \\
i\left[-\alpha_{r}\left(
\partial_{\rho}+\frac{1}{\rho}\right)+\frac{n+1/2}{\rho}\right]  & -\sqrt{E}
\end{array}
\right)%\times \nonumber \\
\left(
\begin{array}{c}
\sqrt{E}u_{n}^{(1)}(\rho)  \\
\sqrt{E}u_{n}^{(2)}(\rho)
\end{array}
\right)
=0.
\end{eqnarray}
The solutions for $E^{2}>0$ are given in terms of the ordinary Bessel functions
\begin{subequations}
\label{bessel}
\begin{eqnarray}
u_{n}^{(1)}(\rho) &=&(\epsilon_{n})^{n} J_{\epsilon_{n}(\nu/\alpha_{r})}(\kappa \rho)   \mbox{,}  \\
u_{n}^{(2)}(\rho) &=&(\epsilon_{n})^{n+1}  J_{\epsilon_{n}(\nu/\alpha_{r} +1)}(\kappa \rho) \mbox{,}
\end{eqnarray}
\end{subequations}
where $\nu\equiv (1-\alpha_{r})/2$, $\kappa=(1/\alpha_{r})\sqrt{E^{2}}$ and $\epsilon_{n}=\pm 1$. The solution (\ref{bessel}) is regular at the origin.

We are interested in the behavior of the solution at large distances from the core region, $k\rho >>1$. The wave function in the asymptotic region is given by
\begin{equation}
\Psi(\rho,\phi) \rightarrow
\sqrt{\frac{i}{k}}f_{k}(\phi)\frac{e^{ik\rho}}{\sqrt{\rho}}\;.
\label{outgoing}
\end{equation}
 The scattered wave function presented here differs from the usual one in two and three dimensions by a factor $\sqrt{i/k}$. As shown in~\cite{adhikari}, this way of defining the scattering amplitude $f_{k}(\phi)$  implies its desirable analytic properties and leads to an optical theorem similar to the three-dimensional one. The  asymptotic form of the Bessel function  is
\begin{subequations}
\label{spinor}
\begin{eqnarray}
\label{spinor1}
u_{n}^{(1)}(\rho) &\approx &(\epsilon_{n})^{n} \sqrt{\frac{2}{\pi \kappa \rho}}
        \cos \left[\kappa \rho -\frac{|\epsilon_{n}\nu|\pi}{2\alpha_{r}}-\frac{\pi}{4}-\frac{(|\epsilon_{n}\nu|-\epsilon_{n}\nu)\pi}{2} \right]   \mbox{,}  \\
\label{spinor2}
u_{n}^{(2)}(\rho) &\approx &(\epsilon_{n})^{n+1}  \sqrt{\frac{2}{\pi \kappa \rho}}
        \cos \left[\kappa \rho -\frac{|\epsilon_{n}\nu+1|\pi}{2\alpha_{r}}-\frac{\pi}{4}-\frac{(|\epsilon_{n}\nu+1|-\epsilon_{n}\nu+1)\pi}{2}  \right] \mbox{.}
\end{eqnarray}
\end{subequations}
From  the asymptotic form of the above spinor (\ref{spinor}), we can obtain the scattering phase shift $\delta_{m}$, which is given by
\begin{equation}
\delta_{m}=\frac{\alpha_{r} -1}{\alpha_{r}}\frac{|m|\pi}{2},
\end{equation}
which can assume  the following values: $m=\epsilon_{n}\nu$ for the first spinor component,  and $m=\epsilon_{n}\nu+1$, for the second one. The property $\delta_{m}=\delta_{-m}$ of the phase shift ensures the absence of back scattering. It follows directly that the  scattering amplitude defined in terms of the phase shift is given by
\begin{eqnarray}
f_{k}(\phi)&=&\sqrt{\frac{2}{\pi}}\sum_{m=-\infty}^{\infty}\sin\delta_{m}e^{i\delta_{m}}e^{im\phi}
\nonumber\\
&=&\sqrt{\frac{2}{\pi}}\sum_{m=-\infty}^{\infty}\sin\left(
\frac{\alpha_{r} -1}{\alpha_{r}}\frac{|m|\pi}{2}  \right) \exp\left(i\;
\frac{\alpha_{r} -1}{\alpha_{r}}\frac{|m|\pi}{2}
 \right)\exp(im\phi).
\end{eqnarray}

\begin{figure}
    \centering
    \includegraphics[scale=1]{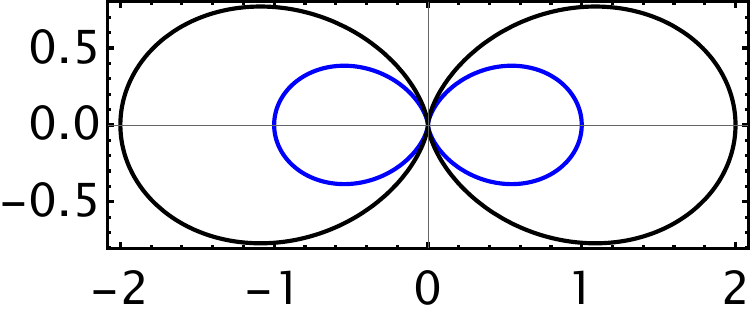}
    \caption{Scattering phase shift for a radial disgyration as a function of the angular variable \(\phi\). The results are shown for different values of the parameter \(\alpha_j\), which characterizes the geometric configuration of the system.}
    \label{phaseshift}
\end{figure}

The scattering phase shift reflects the change in the wave's behavior after interacting with the topological defect. Quantifies the deviation from the original wave behavior due to the scattering event as shown in Fig.\ref{phaseshift}.
This result  allows us to establish the Optical Theorem  relating the imaginary part of the scattering amplitude $f_{k}(\phi)$ to the total cross-section $\lambda$  given by
\begin{equation}
\lambda=\int_{0}^{2\pi\alpha_{r}}\lambda(\phi)d\phi,
\end{equation}
where $\lambda(\phi)$  looks like
\begin{eqnarray}
\lambda(\phi)&=&\frac{|f_{k}(\phi)|^{2}}{k} =\frac{4\alpha_{r}}{k}\sum_{m=-\infty}^{\infty}\sin^{2} \left[
\left( \frac{\alpha_{r}-1}{\alpha_{r}}\right) \frac{|m|\pi}{2}\right].
\end{eqnarray}
The imaginary part of the scattering amplitude  has the form
\begin{equation}
Im f(0)=\sqrt{\frac{2}{\pi}}\sum_{m=-\infty}^{\infty}\sin^{2}
\left[ \left( \frac{\alpha_{r}-1}{\alpha_{r}}\right)
\frac{|m|\pi}{2}\right],
\end{equation}
Therefore the Optical Theorem  can be written as
\begin{equation}
\lambda=\frac{\sqrt{8\pi}\;\alpha_{r}}{k}Im f_{k}(0).
\end{equation}
It is worth noticing that, when $\alpha_{r}=1$ (i.e, no defect, or  the space is Euclidean space),  $f_{k}(0)=0$, as one should expect, since in this case there is no scattering, just the incoming wave.

%%%%%%%%%%%%%%%%%%%%%%%%%%%%%%%%%%%%%
%%%%%%%%%%%%%%%%%%%%%%%%%%%%%%%%%%%%%
\section{Scattering by a Symmetric Vortex}
%%%%%%%%%%%%%%%%%%%%%%%%%%%%%%%%%%%%%
%%%%%%%%%%%%%%%%%%%%%%%%%%%%%%%%%%%%%

 Besides the radial disgyration, we can analyze the fermionic scattering by a symmetric vortex. This vortex can be obtained for the $^3He-A$ in a thin film, where it acts as a superfluid that rotates around the normal vortex axis of the film. For now, we have considered the fermionic quasiparticles far from the vortex axis, where their geometry is represented by the line element \eqref{LEaxi2}. In order to solve the massless Dirac equation \eqref{dirac}, we can define the {\it dreibein} for the symmetric vortex metric:
\begin{subequations}
\begin{eqnarray}
e^{\ \mu}_{0}&=&\delta^{\ \mu}_{0} - \frac{c_{\bot}}{r\omega} \delta^{\ \mu}_{2};\\
e^{\ \mu}_{1}&=& c_{\bot} \delta^{\ \mu}_{1}\\
e^{\ \mu}_{2}&=& \frac{c_{\bot}}{r} \delta^{\ \mu}_{2};\\
e^{\ \mu}_{3}&=& c_{||} \delta^{\ \mu}_{3}
\mbox{.}
\end{eqnarray}
\end{subequations}
And for $(3+1)$-dimensions massless system, the spin connection is given by the Dirac matrices in the Weyl representation. Therefore,
\begin{eqnarray}
    D_{\mu} = \partial_{\mu} + \frac{i}{4} \omega_{\mu ab} \Sigma^{ab},
\end{eqnarray}
with the spin matrice defined through of $\Sigma^{ab} = \frac{i}{2}\left[\gamma^a, \gamma^b\right]$, where the Dirac matrices obeys the anticommutation relation $\lbrace \gamma^{a},\gamma^{b}\rbrace = -2\eta^{ab} I_{4}$. And thus, we can write the massless Dirac equation in this curved background given by the equation \eqref{LEaxi2},
\begin{equation}
    i\frac{\gamma^0}{c_{\bot}} \frac{\partial\psi}{\partial t} + \gamma^{1} \left( \frac{\partial}{\partial \rho} + \frac{1}{2\rho}\right)\psi + i\frac{\gamma^{2}}{r} \left(\frac{\partial}{\partial \phi} - \frac{1}{\omega} \frac{\partial}{\partial t}\right)\psi + i\gamma^{3}\frac{c_{||}}{c_{\bot}} \frac{\partial\psi}{\partial z} = 0.
    \label{wely}
\end{equation}

In order to solve this equation above, we can define the {\it ansatz} in the following way: 
\begin{eqnarray}
    \psi(t, \rho, \phi, z) = e^{-iEt + ij\phi + ikz} \left( \begin{array}{c}
         u_{1\uparrow}(\rho) \\
         u_{1\downarrow}(\rho) \\
         u_{2\uparrow}(\rho)\\
         u_{2\downarrow}(\rho)
    \end{array}\right).
\end{eqnarray}
Substituting this {\it ansatz} in the Weyl equation \eqref{wely}, we can decouple this set of four differential equations. Therefore, we can write the Weyl equation as
\begin{eqnarray}
    \left(\begin{array}{cc}
       \left[\sigma_z\frac{E}{c_{\bot}} + k\frac{c_{||}}{c_{\bot}}\right] & i\left[\frac{d}{d\rho} + \frac{1}{2\rho} - \frac{1}{r}(j+\frac{E}{\omega})\right] \\
       i\left[\frac{d}{d\rho} + \frac{1}{2\rho} + \frac{1}{r}(j+\frac{E}{\omega})\right]  &  -\left[\sigma_z\frac{E}{c_{\bot}} + k\frac{c_{||}}{c_{\bot}}\right]
    \end{array}\right) \left(\begin{array}{cc}
        u_{i\uparrow}  \\
        u_{i\downarrow}
    \end{array}\right) = 0
\end{eqnarray}
In here we have that $\sigma_z = \pm 1$ are the eigenvalues of the pauli matrices. Decoupling this equation, we obtain four decoupled differential equation in a compact way:
\begin{equation}
    \frac{d^{2}u_{is}}{d\rho^{2}} + \frac{1}{r}\frac{du_{is}}{dr} - \Bigg\lbrace\frac{\nu^{2}_{s}}{\rho^{2}} - \frac{E^2}{c^{2}_{\bot}} + k^{2}\frac{c^{2}_{\bot}}{c^{2}_{||}}\Bigg\rbrace u_{is} = 0.
\end{equation}
Performing the coordinate transformation $\rho = \kappa r$, we obtain the Bessel differential equation:
\begin{equation}
    r^{2} \frac{d^2 u_{is}}{dr^2} + r\frac{du_{is}}{dr} + (r^2 - \nu^{2}_{s}) u_{is} = 0,
    \label{bessel2}
\end{equation}
where,
\begin{equation}
    \kappa^{2} = \left(\frac{E^2}{c^{2}_{\bot}} - k^2\frac{c^{2}_{\bot}}{c^{2}_{||}}\right)^{-1}\ \text{and}\ \nu^{2}_{s} = \left(j+\frac{E}{\omega}-\frac{\sigma_z}{2}\right)^{2}.
\end{equation}
The general solution of Eq. \ref{bessel2} can be written as
\begin{eqnarray}
u_{is}(\rho) = A J_{\nu_s}(\kappa \rho) + B Y_{\nu_s}(\kappa \rho),
\end{eqnarray}
where $J_{\nu_s}(\kappa \rho)$ and $Y_{\nu_s}(\kappa \rho)$ are the Bessel functions of the first and second kind, respectively, and $A$ and $B$ are the constants of integration. 

To ensure that the solution is physically meaningful, we impose specific boundary conditions. First, we require regularity at the origin ($\rho \to 0$), which eliminates the term $Y_{\nu_s}(\kappa \rho)$ since it diverges as $\rho \to 0$. Thus, the solution simplifies to:
\begin{eqnarray}
u_{is}(\rho) = A J_{\nu_s}(\kappa \rho).
\end{eqnarray}
Next, we analyze the asymptotic behavior of the solution for $\rho \to \infty$. In this regime, the Bessel function $J_{\nu_s}(\kappa \rho)$ has the following asymptotic form
\begin{eqnarray}
J_{\nu_s}(\kappa \rho) \sim \sqrt{\frac{2}{\pi \kappa \rho}} \cos\left(\kappa \rho - \frac{\nu_s \pi}{2} - \frac{\pi}{4}\right).
\end{eqnarray}
This expression describes an incident wave and a scattered wave, reflecting the physical nature of the scattering process. And then, the general solution satisfying both boundary conditions is writes as
\begin{eqnarray}
u_{is}(\rho) = A J_{\nu_s}(\kappa \rho),
\end{eqnarray}
with its asymptotic behavior for large $\rho$ given by
\begin{eqnarray}
u_{is}(\rho) \sim \sqrt{\frac{2}{\pi \kappa \rho}} \left[ e^{i\left(\kappa \rho - \frac{\nu_s \pi}{2} - \frac{\pi}{4}\right)} + e^{-i\left(\kappa \rho - \frac{\nu_s \pi}{2} - \frac{\pi}{4}\right)} \right].
\end{eqnarray}

\begin{figure}
    \centering
    \includegraphics[scale=1]{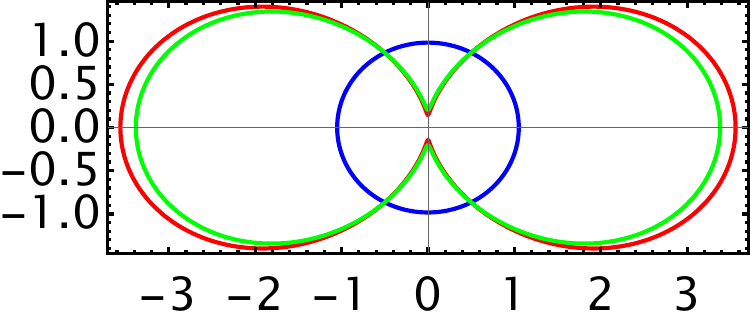}
    \caption{Scattering phase shift for a symmetric vortex as a function of the angular variable \(\phi\), considering the summation over the quantum modes \(\nu_s\). The results are shown for different sets of parameters \((E, \omega, \sigma_z)\), which characterize the energy, rotation, and spin configurations of the system.}
    \label{phaseshift2}
\end{figure}

This form naturally separates the solution into components that represent the incident wave and the scattered wave. The scattered wave can be further analyzed to determine the scattering amplitude and the total cross section. The scattered wave is represented by the outgoing term Eq. (\ref{outgoing}).  By comparing this expression with the general solution, we can identify $f(\phi)$ in terms of the scattering phase shift
\begin{eqnarray}
f(\phi) = \sum_{\nu_s} \frac{1}{\sqrt{k}} e^{i \delta_{\nu_s}} \sin(\delta_{\nu_s}) e^{i \nu_s \phi},
\end{eqnarray}
For symmetric vortices, the phase shift is given by:
\begin{eqnarray}
\delta_{\nu_s} = -\frac{\pi}{2} \left( \frac{E}{\omega} - \frac{\sigma_z}{2} - \frac{1}{2} \right).
\end{eqnarray}
This expression reflects the complete dependence of the scattering phase $\delta_{\nu_s}$ on the index $\nu_s$, which is determined by the system dynamics as seen in Fig.\ref{phaseshift2}. As we have seen before, the total cross section $\sigma_{\text{total}}$ is obtained by integrating the squared modulus of the scattering amplitude over all angles. Therefore, the total cross section becomes
\begin{eqnarray}
\lambda= \frac{8\pi}{\kappa} \sum_{\nu_s=-\infty}^\infty \sin^2(\delta_{\nu_s}).
\end{eqnarray}

The optical theorem, originally studied for radial disgyrations, extends naturally to symmetric vortices, linking the imaginary part of the forward scattering amplitude to the total cross-section. For symmetric vortices, this relationship retains the same formal structure but incorporates the unique features of the vortex geometry, such as rotational dynamics and their influence on the phase shifts. By representing the forward-scattering amplitude through the phase shifts, the total cross section emerges as a summation over all angular momenta, reflecting the interaction between the vortex's rotational properties and the scattered wave. This validates the optical theorem in the context of symmetric vortices without requiring modifications to its general framework. The application of the optical theorem to symmetric vortices complements its derivation for radial disgyrations, demonstrating the approach's versatility across different topological and geometrical configurations. This result emphasizes that the total cross section encapsulates essential physical properties of the system, offering a unified description of scattering phenomena in superfluids.

%%%%%%%%%%%%%%%%%%%%%%%%%%%%%%%%%%%%%%%%%%%%%%%%%%%%%%%%%%%%%%%%%%%%%%%%%%%%%%%%%%%%
\section{Concluding Remarks}

In this work, we used the analog model developed by Volovik to investigate the quantum dynamics of massless quasiparticles in superfluid \( ^3\text{He}-A \)~\cite{volo,boo} in the presence of topological defects. We focus on two distinct cases: the scattering of quasiparticles by radial disgyrations and symmetric vortices. These systems provide an intriguing platform for exploring the interplay between geometry, topology, and quantum dynamics, offering analogs to gravitational phenomena in condensed matter systems.

For the radial disgyration, we analyzed the effects of the conical geometry induced by the defect, characterized by the parameter \(\alpha_r\), on the scattering behavior. This analysis allowed us to derive the scattering amplitude, the total cross-section, and the corresponding phase shifts. Additionally, we verified the optical theorem in this context, establishing the relationship between the imaginary part of the forward scattering amplitude and the total cross section. The results revealed a high sensitivity of the scattering properties to the geometric parameter \(\alpha_r\), providing insight into how the defect topology governs the scattering dynamics.

In the case of symmetric vortices, we extended the analysis to include the rotational dynamics of the superfluid, represented by the angular velocity \(\omega\). The scattering amplitude and phase shifts were derived for this case, demonstrating how the rotational properties of the vortex influence the scattering process. Despite the added complexity of the vortex geometry, the optical theorem was shown to hold, highlighting the robustness of the theoretical framework across different topological defects. This case further underscored the role of angular momentum and spin in shaping the scattering behavior.

The combined analysis of these two cases illustrates the versatility of the analog model in capturing a wide range of physical effects induced by topological defects in quantum fluids. The formalism used not only provides a unified description of scattering phenomena but also serves as a bridge between theoretical predictions and potential experimental realizations. In particular, the dependence of the scattering properties on the geometric and dynamical parameters suggests the possibility of experimentally probing these effects in superfluid systems, thereby validating the theoretical results.

In conclusion, our work expands the understanding of quasiparticle dynamics in superfluid \( ^3\text{He}-A \), emphasizing the profound influence of the geometry and topology of the defect on scattering phenomena. The verification of the optical theorem for both radial disgyrations and symmetric vortices demonstrates the consistency and applicability of the analog model across different defect configurations. These findings open new avenues for experimental exploration and deepen the connection between condensed matter systems and analog gravitational models.
%%%%%%%%%%%%%%%%%%%%%%%%%%%%%%%%%%%%%%%%%%%%%%%%%%%%%%%%%%%%%%%%%%%%%%%%%%%%%%%
%%%%%%%%%%%%%%%%%%%%%%%%%%%%%%%%%%%%%%%%%%%%%%%%%%%%%%%%%%%%%%%%%%%%%%%%%%%%%%%

{\bf Acknowledgements.} We {\bf thank} CAPES, CNPq 302485/2023-6 and CAPES/NANOBIOTEC  for financial support. G. Q. Garcia would like to thank Fapesq-PB for financial support (Grant BLD-ADT-A2377/2024). The work of C. Furtado has been supported by the CNPq (project PQ Grant 1A No. 311781/2021-7).
%%%%%%%%%%%%%%%%%%%%%%%%%%%%%%%%%%%%%%%%%%%%%%%%%%%%%%%%%%%%%%%%%%%%%%%%%%%%%%%


\begin{thebibliography}{99}
\bibitem{scie:bow} M. J. Bowick et al., Science {\bf 263}, 943 (1994).

\bibitem{pr:zur} W. H. Zurek, Phys. Rep. {\bf 276}, 177 (1996).

\bibitem{prl:gar} L. J. Garay, J. R. Anglin, J. I. Cirac and P. Zoller, Phys. Rev. Lett. {\bf 85}, 4643 (2000).

\bibitem{pra:gar} L. J. Garay, J. R. Anglin, J. I. Cirac and P. Zoller, Phys. Rev. A {\bf 63}, 023611 (2001).

\bibitem{prl:unruh} W. G. Unruh, Phys. Rev. D {\bf 51}, 2827 (1995).

\bibitem{prd:unruh} W. G. Unruh, Phys. Rev. D {\bf 51}, 6 (1995).

\bibitem{prd:jac} T. Jacobson, Phys. Rev. D {\bf 44}, 1731 (1991).

\bibitem{prl:visser} M. Visser, Phys. Rev. Lett. {\bf 80}, 3436 (1998).

\bibitem{cqg:visser} M. Visser, Class. Quantum Grav. {\bf 15}, 1767 (1998).

\bibitem{volo} G. E. Volovik, JETP Lett. {\bf 67}, 11 (1998).

\bibitem{garciadis} L. C. Garcia de Andrade, A. M. de M. Carvalho and C. Furtado, Europhys. Lett. {\bf 67}, 538 (2004).

\bibitem{Sta} D. D. Sokolov and A. A. Starobinskii, Sov. Phys. Dokl. {\bf 22}, 312 (1977).

\bibitem{Vil1} A. Vilenkin, Phys. Rep. {\bf 121}, 263 (1985).

\bibitem{His} A. Vilenkin, Phys. Lett. B {\bf 133}, 177 (1983); W. A. Hiscock, Phys. Rev. A {\bf 31}, 3288 (1985); B. Linet, Gen. Rel. Grav. {\bf 17}, 1109 (1985).

\bibitem{Bar} M. Barriola and A. Vilenkin, Phys. Rev. Lett. {\bf 63}, 341 (1989).

\bibitem{Vol} R. H. Brandenberger, Physica B {\bf 178}, 42 (1992).

\bibitem{Kat} M. O. Katanaev and I. V. Volovich, Ann. Phys. (NY) {\bf 216}, 1 (1992); C. Furtado and F. Moraes, Phys. Lett. A {\bf 188}, 394 (1994).

\bibitem{Kib} T. W. Kibble, J. Phys. A {\bf 9}, 1387 (1976); Y. B. Zeldovich, Mon. Not. R. Astron. Soc. {\bf 192}, 663 (1980).

\bibitem{boo} G. E. Volovik, {\it The Universe in a Helium Droplet} (Oxford University Press, 2003).

\bibitem{leon} U. Leonhardt, Phys. Rev. A {\bf 65}, 043818 (2002).

\bibitem{brev} I. Brevik et al., Phys. Rev. D {\bf 65}, 024005 (2002).

\bibitem{prd:novello} M. Novello, Int. J. Mod. Phys. A {\bf 17}, 4187 (2002).

\bibitem{eduardo} M. A. Anacleto, F. A. Brito and E. Passos, Phys. Rev. D {\bf 85}, 025013 (2012).

\bibitem{anacleto1} M. A. Anacleto, F. A. Brito and E. Passos, Phys. Lett. B {\bf 694}, 149 (2010).

\bibitem{anacleto2} M. A. Anacleto, F. A. Brito and E. Passos, Phys. Lett. B {\bf 703}, 609 (2011).

\bibitem{anacleto3} M. A. Anacleto, F. A. Brito and E. Passos, Phys. Rev. D {\bf 87}, 125015 (2013).

\bibitem{3} G. E. Volovik, Low Temp. Phys. {\bf 24}, 2 (1998).

\bibitem{staro} D. D. Sokolov and A. A. Starobinskii, Sov. Phys. Dokl. {\bf 22}, 312 (1977).

\bibitem{Kro} E. Kr\"oner, Continuum theory of defects, in *Les Houches, Session XXXV, 1980 - Physics of Defects*, eds. R. Balian et al. (North-Holland, Amsterdam, 1981), pp. 282-315.

\bibitem{kat} M. Katanaev and I. Volovich, Ann. Phys. {\bf 216}, 1 (1992).

\bibitem{adhikari} S. K. Adhikari, Am. J. Phys. {\bf 54}, 362 (1986).

\bibitem{azevedo} S. Azevedo and F. Moraes, J. Phys.: Condens. Matter {\bf 12}, 7421 (2000).

\bibitem{exotic} G. E. Volovik, {\it Exotic Properties of Superfluid He-3} (World Scientific, Singapore, 1992).




%%%%%%%%%%%%%%%%%%%%%%%%%%%%%%%%
%%%%%%%%%%%%%%%%%%%%%%%%%%%

\end{thebibliography}
\end{document}